\documentclass[aps,prx,reprint,superscriptaddress]{revtex4-1}
\usepackage{amsmath,amssymb,bm}

\begin{document}

\title{Comment on ``Extension of the adiabatic theorem''}

\author{Jie Gu}
\affiliation{Chengdu Academy of Educational Sciences, Chengdu 610036, China}

\date{\today}

\begin{abstract}
In Ref.~\cite{DamerowKehrein2026}, Damerow and Kehrein proposed the conjecture
that, for quantum quenches within the same phase, the overlap between the initial
ground state and postquench eigenstates is maximal for the postquench ground state.
We show that this statement is not valid in general. An explicit local,
translationally invariant, gapped free-fermion counterexample exists even though
the pre- and postquench Hamiltonians are connected by a symmetry-preserving gapped
path and the thermodynamic-limit spectrum is continuous.
\end{abstract}

\maketitle

In Ref.~\cite{DamerowKehrein2026}, the following conjecture was formulated
[Eq.~(2) of that work]:
\begin{equation}
\max_n \left| \langle \psi_n^{(f)} | GS_i \rangle \right|^2
=
\left| \langle GS_f | GS_i \rangle \right|^2 ,
\label{eq:conjecture}
\end{equation}
for quenches between Hamiltonians in the same phase.
Here $|GS_i\rangle$ and $|GS_f\rangle$ denote the pre- and postquench ground states,
respectively, and $\{|\psi_n^{(f)}\rangle\}$ are the eigenstates of the postquench
Hamiltonian.
We provide a counterexample.

Consider a one-dimensional translationally invariant two-band free-fermion model
\begin{equation}
H_\mu = \sum_k \Psi_k^\dagger h_\mu(k)\Psi_k,
\qquad
\Psi_k = \begin{pmatrix} a_k \\ b_k \end{pmatrix},
\end{equation}
with
\begin{subequations}
\begin{align}
h_f(k) &= \varepsilon(k)\,\sigma_z, \\
h_i(k) &= \varepsilon(k)\,\bigl(\cos\phi\,\sigma_z + \sin\phi\,\sigma_x \bigr),
\end{align}
\end{subequations}
and
\begin{equation}
\varepsilon(k)=m+t\cos k,
\qquad
m>|t|>0.
\label{eq:dispersion}
\end{equation}
Because $\varepsilon(k)$ contains only a constant and a cosine harmonic, both
Hamiltonians are local nearest-neighbor models in real space.

At half filling, both $H_i$ and $H_f$ are gapped. Their single-particle eigenvalues are
\begin{equation}
E_\pm(k)=\pm \varepsilon(k),
\end{equation}
so the gap is finite for all $k$. Moreover, the interpolation
\begin{equation}
h_s(k)=\varepsilon(k)\,\bigl[\cos(s\phi)\,\sigma_z+\sin(s\phi)\,\sigma_x\bigr],
\qquad s\in[0,1],
\label{eq:path}
\end{equation}
preserves translation symmetry and particle-number conservation, while its spectrum
remains $\pm \varepsilon(k)$ for all $s$. Hence $H_i$ and $H_f$ lie in the same phase
according to the definition adopted in Ref.~\cite{DamerowKehrein2026}. In the
thermodynamic limit, $k$ becomes continuous, and the spectrum is correspondingly
continuous as required by the premise of Ref.~\cite{DamerowKehrein2026}.

Let $\beta_{k,\pm}^\dagger$ create the negative- and positive-energy eigenstates of
$h_f(k)$. The occupied negative-energy state of $h_i(k)$ is then
\begin{equation}
d_k^\dagger
=
\cos\frac{\phi}{2}\,\beta_{k,-}^\dagger
-
\sin\frac{\phi}{2}\,\beta_{k,+}^\dagger .
\label{eq:rotation}
\end{equation}
For a finite system with $N$ occupied momenta, the initial ground state is
\begin{equation}
|GS_i\rangle = \prod_k d_k^\dagger |0\rangle ,
\end{equation}
whereas the postquench ground state is
\begin{equation}
|GS_f\rangle = \prod_k \beta_{k,-}^\dagger |0\rangle .
\end{equation}

Any postquench eigenstate at the same filling is specified by a subset $S$ of momenta
promoted to the upper band:
\begin{equation}
|S\rangle
=
\prod_{k\in S}\beta_{k,+}^\dagger
\prod_{k\notin S}\beta_{k,-}^\dagger
|0\rangle .
\end{equation}
Since different momentum sectors factorize independently, one immediately finds
\begin{equation}
\left| \langle S|GS_i\rangle \right|^2
=
\left(\cos^2\frac{\phi}{2}\right)^{N-|S|}
\left(\sin^2\frac{\phi}{2}\right)^{|S|}.
\label{eq:generaloverlap}
\end{equation}
In particular, for the postquench ground state,
\begin{equation}
\left| \langle GS_f|GS_i\rangle \right|^2
=
\left(\cos^2\frac{\phi}{2}\right)^N .
\label{eq:gsoverlap}
\end{equation}

Now consider a single particle-hole excitation $|q\rangle$, corresponding to $|S|=1$.
Equation~(\ref{eq:generaloverlap}) gives
\begin{equation}
\frac{\left| \langle q|GS_i\rangle \right|^2}
{\left| \langle GS_f|GS_i\rangle \right|^2}
=
\tan^2\frac{\phi}{2}.
\label{eq:singleexcitation}
\end{equation}
Therefore, whenever $\phi>\pi/2$, one has
\begin{equation}
\left| \langle q|GS_i\rangle \right|^2
>
\left| \langle GS_f|GS_i\rangle \right|^2,
\end{equation}
so the conjecture in Eq.~(\ref{eq:conjecture}) is violated.

For a concrete choice, take
\begin{equation}
\phi=\frac{3\pi}{4}.
\end{equation}
Then
\begin{equation}
\tan^2\frac{\phi}{2}
=
\tan^2\frac{3\pi}{8}
=
(1+\sqrt{2})^2
>
1,
\end{equation}
and thus even a single excited eigenstate has larger overlap with the initial ground
state than the final ground state.

In fact, Eq.~(\ref{eq:generaloverlap}) shows more. If $\phi>\pi/2$, then
\begin{equation}
\sin^2\frac{\phi}{2}>\cos^2\frac{\phi}{2},
\end{equation}
so the overlap increases monotonically with $|S|$. The maximal overlap is therefore
attained for the highly excited state in which all occupied momenta are promoted to the
upper band:
\begin{equation}
|S_{\max}\rangle = \prod_k \beta_{k,+}^\dagger |0\rangle ,
\end{equation}
for which
\begin{equation}
\left| \langle S_{\max}|GS_i\rangle \right|^2
=
\left(\sin^2\frac{\phi}{2}\right)^N
>
\left(\cos^2\frac{\phi}{2}\right)^N
=
\left| \langle GS_f|GS_i\rangle \right|^2.
\end{equation}

We conclude that the conjecture proposed in Ref.~\cite{DamerowKehrein2026} does not
hold in general, even within the class of local, translationally invariant, gapped
free-fermion Hamiltonians connected by a symmetry-preserving gapped path and having a
continuous spectrum in the thermodynamic limit. The positive results obtained in
Ref.~\cite{DamerowKehrein2026} for the TFIM and special cases of the ANNNI model must
therefore rely on additional structure not captured by the conjecture in its present
form.

\emph{Note added: Subsequent to submission, we note a preprint deriving an exact necessary-and-sufficient criterion for identifying when ground-state dominance is valid or violated for translational-invariant free-fermion Hamiltonians \cite{Haque2026}, providing independent support for our counterexample.}

\end{document}